\documentclass{article}
\usepackage{amsmath}
\usepackage{amssymb}
\usepackage{bm}
\usepackage{url}
\begin{document}
\begin{center}
\LARGE{Oscillating Flavors in Massless Neutrinos}
\end{center}
\medskip
\begin{center}
Lester C. Welch

Aiken, SC 29803

lester.welch@gmail.com
\end{center}
\begin{quote}
By considering Dirac's equation using quaternions ($\mathbb{H}$) with their greater degree of freedom in imaginaries, it is shown that a model can be created with oscillations among flavors, even if the particles are massless.  Furthermore the solutions are spin $\frac{1}{2}$ and have helicities depending on whether their energy is positive or negative.
\end{quote}

\section{Introduction}

To give a short review to set the context, using only complex numbers ($\mathbb{C}$) and following Schiff's \cite{Schiff.55} notation and representation, Dirac's equation for a massless particle is ($\hbar=c=1$):
\begin{equation} 
\bigg( i\partial_t-i\boldsymbol{\alpha_c} \cdot\boldsymbol{\bigtriangledown}\bigg)|\psi>=0 \label{eqn:dirac}
\end{equation}
where the elements of $\boldsymbol{\alpha}_c \in \mathbb{C}$ and
\begin{equation*}
\boldsymbol{\bigtriangledown}=\partial_x+\partial_y+\partial_z
\end{equation*} 
One representation for $\boldsymbol{\alpha}_c$ is given in Appendix $\mathbf{A}$.
Equation ($\ref{eqn:dirac}$) has four independent solutions corresponding to positive and negative energy having spin up or spin down:
\begin{equation}
|\psi_{+-}^{\pm}> = u_{+-}^{\pm}\exp^{{i(\mathbf{p} \cdot \mathbf{r} -E t)}} \label{eqn:pw}
\end{equation}
where, for example, 
\begin{equation} 
u_{+}^{+} = \frac{1}{\sqrt{2}}\left(\begin{array}{c}
-\frac{p_z}{E}\\
 \\
-\frac{(p_x+ip_y)}{E}\\
 \\
1\\
 \\
0\\
\end{array}\right) \label{eqn:sol}
\end{equation}
Note, in general, that $u^{-1}= (u^*)^T$.

\section{Quaternion Solution}
Welch \cite{welch1} has explored Dirac's equation for massive charged particles and showed that charges $\frac{2}{3}$ and $\frac{1}{3}$ result with a minimum of assumptions using quaternions (see Appendix $\mathbf{B}$).  
Defining $q$ to be a constant unit imaginary quaternion (see Appendix $\mathbf{B}$)
the transition to quaternions could trivially and unproductively follow the complex analysis exactly because any unit imaginary quaternion, $q_i$, is isomorphic to $i \in \mathbb{C}$.  Equation ($\ref{eqn:pw}$) 
becomes 
\begin{equation*}
|\psi_q>\equiv|\psi_{q+-}^{\pm}>= u_{q+-}^{\pm}\exp^{{q_i(\mathbf{p} \cdot \mathbf{r} -E t)}}
\end{equation*} 
where, for example, equation($\ref{eqn:sol}$) becomes
\begin{equation*} 
u_{q+}^{+} = \frac{1}{\sqrt{2}}\left(\begin{array}{cccc}
-\frac{p_z}{E}\\
 \\
-\frac{(p_x+q_ip_y)}{E}\\
 \\
1\\
 \\
0\\
\end{array}\right) 
\end{equation*} \\

However, we wish to explore the ramifications possible due to the increased flexibility of quaternions, thus repeating the derivation in $\mathbb{H}$ that is done in  $\mathbb{C}$ with equation(\ref{eqn:dirac}) in the form, in which no imaginary unit from $\in \mathbb{C}$ appears
\begin{equation} 
\bigg( \partial_t-\boldsymbol{\alpha} \cdot\boldsymbol{\bigtriangledown}\bigg)|\psi_q>=0
 \label{eqn:diracq}
\end{equation}
with the elements of $\boldsymbol{\alpha} \in \mathbb{H}$.
Requiring that $|\psi_q>$ also be a solution to Klein-Gordon's relativistic equation for a massless particle, i.e., 
\begin{equation*}
(\partial^2_t- \boldsymbol{\bigtriangledown}^2)|\psi_q>=0 
\end{equation*} leads to the necessity that $\boldsymbol{\alpha}$ be a $2$ x $2$ matrix and obey the commutative relationships:
\begin{eqnarray*}
\alpha_x^2=\alpha_y^2=\alpha_z^2&=&1\\
\alpha_x\alpha_y+\alpha_y\alpha_x=\alpha_y\alpha_z+\alpha_z\alpha_y=\alpha_z\alpha_x+\alpha_x\alpha_z&=&0\\
\end{eqnarray*}
One such quaternion representation of Dirac's $\boldsymbol{\alpha}$  is given in Appendix $\mathbf{A}$.
Thus equation ($\ref{eqn:diracq}$) is
\begin{equation*}
\left( \begin{array}{cc}
 \partial_t & 0\\
 0 & \partial_t
\end{array} \right)
-\left( \begin{array}{cc}
 0 & i\partial_x\\
 -i\partial_x & 0
\end{array} \right) 
-\left( \begin{array}{cc}
 0 & j\partial_y\\
 -j\partial_y & 0
\end{array} \right) 
-\left( \begin{array}{cc}
 0 & k\partial_z\\
 -k\partial_z & 0
\end{array} \right) 
\left( \begin{array}{c}
 |\psi_{q1}>\\
 |\psi_{q2}>
\end{array} \right) =0
\end{equation*} 
or
\begin{equation}
\left( \begin{array}{cc}
 \partial_t & -i\partial_x-j\partial_y-k\partial_z\\
 i\partial_x+j\partial_y+k\partial_z & \partial_t
\end{array} \right) 
\left( \begin{array}{c}
 |\psi_{q1}>\\
 |\psi_{q2}>
\end{array} \right)=0 \label{eqn:solq1}
\end{equation}
Let us use as a trial solution:\footnote{Given the non-commutative nature of quaternions, an equally valid trial solution, not pursued herein, is
\begin{equation*}
|\psi_q> = \exp^{q(\mathbf{p} \cdot \mathbf{r} -E t)}u^R
\end{equation*}}
\begin{equation*}
|\psi_q> = u\exp^{q(\mathbf{p} \cdot \mathbf{r} -E t))}
\end{equation*}
Thus equation (\ref{eqn:solq1}) can be written:
\begin{equation}
\left( \begin{array}{cc}
 \partial_t & -i\partial_x-j\partial_y-k\partial_z\\
 i\partial_x+j\partial_y+k\partial_z & \partial_t\\
\end{array} \right)\\ 
\left( \begin{array}{c}
 u_1\\
 u_2\\
\end{array}\right) \exp^{q(\mathbf{p} \cdot \mathbf{r} -E t)}=0 \label{eqn:hal}\\
\end{equation}
which has the general solutions (see Appendix $\mathbf{C}$).
\begin{eqnarray*}
u_2^*u_2&=&\frac{1}{2} \\
u_1 &=&\frac{-u_2 (\mathbf{\epsilon} \cdot \mathbf{p})}{E} \\
\end{eqnarray*}
Note
\begin{equation*}
u_1^*u_2 = \frac{(\mathbf{\epsilon} \cdot \mathbf{p})}{E}u_2^*u_2=\frac{(\mathbf{\epsilon} \cdot \mathbf{p})}{2E}
\end{equation*}
If the "spin up" state is the positive energy solution then the "spin down" state is the negative energy solution - i.e., the anti-neutrino, as is seen experimentally.  Having only two spin states is indicative of a spin $\frac{1}{2}$ system.
Thus the orthonormal solutions are  
\begin{eqnarray*}
u_{+}= \frac{1}{\sqrt{2}}\left( \begin{array}{c}
-\frac{(\mathbf{\epsilon} \cdot \mathbf{p})}{E}\\
 \\
 1 \\
\end{array}\right)& &
u_{-}= \frac{1}{\sqrt{2}}\left( \begin{array}{c}
1\\
 \\
 -\frac{(\mathbf{\epsilon} \cdot \mathbf{p})}{|E|}\\
\end{array}\right) \\
\end{eqnarray*}
Thus
\begin{equation*}
|\psi_q>=N_{+} \frac{1}{\sqrt{2}}  \left( \begin{array}{c}
 -\frac{(\mathbf{\epsilon} \cdot \mathbf{p})}{E}\\
 \\
 1\\
\end{array}\right)\exp^{q(\mathbf{p} \cdot \mathbf{r} -E t)}+N_{-}\frac{1}{\sqrt{2}}   \left( \begin{array}{c}
 1\\
 \\
 -\frac{(\mathbf{\epsilon} \cdot \mathbf{p})}{  |E|}\\
\end{array}\right)\exp^{q(\mathbf{p} \cdot \mathbf{r} -E t)} \\
\end{equation*}
$|\psi_q>$, which has no specific lepton flavor and the most general solution, will be denoted as the "q-neutrino" and is the eigenstate of equation(\ref{eqn:hal}).
and can be written
\begin{equation*}
|\psi_q>=N_{+}|\psi_{q+}> + N_{-}|\psi_{q-}>
\end{equation*}
where $|\psi_{q+}>$ and $|\psi_{q-}>$ are the basis vectors of the Hamiltonian as defined by equation ($\ref{eqn:diracq}$).
The normalizing factors $N_{+}$ and $N_{-}$ obey
\begin{equation*}
N_{+}^2 + N_{-}^2 =1
\end{equation*} and are chosen to $\in \mathbb{R}$.  It is straight forward to show:
\begin{eqnarray*}
<\psi_{q+}|\psi_{q+}>=<\psi_{q-}|\psi_{q-}>&=&1\\
<\psi_{q-}|\psi_{q+}>=<\psi_{q+}|\psi_{q-}>&=&0\\ \\
\end{eqnarray*}
\section{Lepton Flavors}
Looking at the positive energy state,
\begin{eqnarray*}
|\psi_{q+}>&=&\frac{1}{\sqrt{2}}
\left( \begin{array}{c}
 - \frac{1}{E}(\mathbf{\epsilon} \cdot \mathbf{p})\\
 \\
 1\\
\end{array}\right) \exp^{q(\mathbf{p} \cdot \mathbf{r} -E t)}\\ 
\end{eqnarray*}
One can expand $|\psi_{q+}>$ using the basis of $\mathbb{H}$ to obtain
\begin{eqnarray}
|\psi_{q+}>&=&\frac{1}{\sqrt{2}}
\left( \begin{array}{c}
 \frac{1}{E}(ap_x+bp_y+cp_z)\sin{ (\mathbf{p} \cdot \mathbf{r} -E t)}\\
 \\
\cos{(\mathbf{p} \cdot \mathbf{r} -E t)}\\
\end{array}\right)  \nonumber\\ \nonumber \\ \nonumber \\
&+& \frac{i}{\sqrt{2}}
\left( \begin{array}{c}
 -\frac{p_x}{E}\cos{(\mathbf{p} \cdot \mathbf{r} -E t)}-\frac{(cp_y-bp_z)}{E}\sin{(\mathbf{p} \cdot \mathbf{r} -E t)} \\
 \\
 a\sin{(\mathbf{p} \cdot \mathbf{r} -E t)}
\end{array}\right)  \nonumber\\ \nonumber \\  \nonumber\\
&+& \frac{j}{\sqrt{2}}
\left( \begin{array}{c}
 -\frac{p_y}{E}\cos{(\mathbf{p} \cdot \mathbf{r} -E t)}-\frac{(ap_z-cp_x)}{E}\sin{(\mathbf{p} \cdot \mathbf{r} -E t)} \\
 \\
 b\sin{(\mathbf{p} \cdot \mathbf{r} -E t)}
\end{array}\right)  \nonumber\\ \nonumber \\ \nonumber \\
&+& \frac{k}{\sqrt{2}}
\left( \begin{array}{c}
 -\frac{p_z}{E}\cos{(\mathbf{p} \cdot \mathbf{r} -E t)}-\frac{(bp_x-ap_y)}{E}\sin{(\mathbf{p} \cdot \mathbf{r} -E t)} \\
 \\
 c\sin{(\mathbf{p} \cdot \mathbf{r} -E t)}
\end{array}\right) \label{eqn:Hcomp}\\ \nonumber \\  \nonumber \\  \nonumber
\end{eqnarray}
or
\begin{eqnarray*}
|\psi_{q+}>&=&|r_+>+i|e_+>+j|\mu_+>+k|\tau_{+}>\\
\end{eqnarray*}
where an association has been explicitly been made between the lepton flavors and the imaginary bases of quaternions and similarly for $|\psi_{q-}>$\\ \\
Note that this equation is invariant under the simultaneous permutations:
\begin{eqnarray*}
i \rightarrow j \rightarrow k \rightarrow i\\
p_x \rightarrow p_y \rightarrow p_z \rightarrow p_x\\
a \rightarrow b \rightarrow c \rightarrow a\\
\end{eqnarray*}
Thus there is no preferred spatial direction.
\section{Inner Product}
Before we explore the ramifications of this formulation the issue of the definition of "inner product" of Hilbert vectors $A$ and $B$ ($=A \cdot B$) must be addressed.  One mathematical requirement of an "inner product" is that it is commutative, i.e., 
\begin{equation}
A\cdot B= B\cdot A \label{eqn:dotc}
\end{equation}
This clearly is not a problem in $ \mathbb{C}$, a commutative algebra, however in $ \mathbb{H}$, a non-commutative algebra, more care has to be taken since, in general, $A \cdot B \ne B \cdot A$ . The definition used herein is adopted from that used in the special Jordan algebras, i.e., 
\begin{equation}
A\circ B=\frac{1}{2}(A\cdot B+B\cdot A).  \label{eqn:doth}
\end{equation}
Clearly in $ \mathbb{C}$ equation ($\ref{eqn:doth}$) reduces to equation ($\ref{eqn:dotc}$) and thus ordinary quantum mechanics would not be affected by such a generalization.  We will extent this change of definition to the bra-key notation, i.e., the $<A|$ and $|B>$ notation. The bra, $<A|$, is the Hermitian conjugate of the ket $|A>$. So, by definition, 
\begin{equation*}
<B| \circ|A>=\frac{1}{2}( <B|A>+<A|B>)
\end{equation*}
Using this generalization, it is easy to see from equation (\ref{eqn:Hcomp}) that
\begin{eqnarray*}
<r| \circ |r>&=&\frac{1}{2}\bigg(\frac{(ap_x+bp_y+cp_z)^2}{E^2} \sin^2{(\mathbf{p} \cdot \mathbf{r} -E t)}+ \cos^2({(\mathbf{p} \cdot \mathbf{r} -E t)}\bigg)>0\\
<r| \circ |e>&=&<r| \circ |\mu>=<r| \circ |\tau>=0\\
<e| \circ |e>&=&\frac{1}{2}\bigg[\bigg(\frac{p_x}{E} \cos{(\mathbf{p} \cdot \mathbf{r} -E t)}+ \frac{(cp_y-bp_z)}{E}\sin({(\mathbf{p} \cdot \mathbf{r} -E t)}\bigg)^2+a^2 \sin^2{(\mathbf{p} \cdot \mathbf{r} -E t)}\bigg]>0\\
<e| \circ |\mu>&=&<e | \circ |\tau>=<e| \circ|r>=0\\
<\mu| \circ |\mu>&=&\frac{1}{2}\bigg[\bigg(\frac{p_y}{E} \cos{(\mathbf{p} \cdot \mathbf{r} -E t)}+ \frac{(ap_z-cp_x)}{E}\sin({(\mathbf{p} \cdot \mathbf{r} -E t)}\bigg)^2+b^2 \sin^2{(\mathbf{p} \cdot \mathbf{r} -E t)}\bigg]>0\\
<\mu| \circ|\tau>&=&<\mu| \circ |e>=<\mu| \circ |r>=0\\
<\tau| \circ |\tau>&=&\frac{1}{2}\bigg[\bigg(\frac{p_z}{E} \cos{(\mathbf{p} \cdot \mathbf{r} -E t)}+ \frac{(bp_x-ap_y)}{E}\sin({(\mathbf{p} \cdot \mathbf{r} -E t)}\bigg)^2+c^2 \sin^2{(\mathbf{p} \cdot \mathbf{r} -E t)}\bigg]>0\\
\end{eqnarray*}\\ \\
\section{Flavor Oscillation}
Oscillation among lepton flavors is well established \cite{SuK} and the only mechanism theoretically possible in $\mathbb{C}$ is if the neutrinos have mass. A plane wave, for a free particle has the form  ($i \in \mathbb{C}$)
\begin{equation*}
|\psi> = \exp^{i(\vec{p}\cdot \vec{x}-Et)}
\end{equation*}
So for a given direction, the only distinguishing parameter of different plane waves is the energy, $E$.  However, in $\mathbb{H}$, one can have distinct plane waves with the same energy, i.e.,
\begin{eqnarray*}
|\psi_1> = \exp^{q_1(\vec{p}\cdot \vec{x}-Et)}\\
|\psi_2> = \exp^{q_2(\vec{p}\cdot \vec{x}-Et)}\\
\end{eqnarray*}
where $q_1$ and $q_2$ are differing unit imaginary quaternions and ($i,j,k \in \mathbb{H}$),
\begin{eqnarray*}
q_1 =a_1i+b_1j+c_1k & &\\
q_2 =a_2i+b_2j+c_2k& &
\end{eqnarray*}
and
\begin{equation*}
q_1*q_1 = q_2*q_2 = -1\\ 
\end{equation*}
\subsection{In $\mathbb{C}$ }
To illustrate simply the consequences of this difference between $\mathbb{C}$ and $\mathbb{H}$, we follow the example of  Casper \cite{cosc}, a member of the "Super-Kamiokande" Collaboration \cite{SuK}, who derives the requirement of a mass difference within $\mathbb{C}$.\\ \\
An outline of the argument is: 
Let $|\nu_1>$ and $|\nu_2>$ be neutrino eigenstates of mass, $m_1$ and $m_2$, respectively.  The time evolution, for a free particle, by Schrodinger's equation is:
\begin{eqnarray*}
\left(\begin{array}{c}
|\nu_1(\vec{x}, t)>\\
|\nu_2(\vec{x}, t)>\\
\end{array}\right)&=&\exp^{i\vec{p}\cdot \vec{x}}\left(\begin{array}{c}
\exp^{-iE_1t}|\nu_1(0, 0)>\\
\exp^{-iE_2t}|\nu_2(0, 0)>\\
\end{array}\right)\\
&=&\exp^{i\vec{p}\cdot \vec{x}}\left(\begin{array}{cc}
\exp^{-iE_1t}&0\\
0& \exp^{-iE_2t}
\end{array}\right)
\left(\begin{array}{c}
|\nu_1(0, 0)>\\
|\nu_2(0, 0)>\\
\end{array}\right)\\\\
\end{eqnarray*}
The mass eigenstates can also be expressed in terms of flavor eigenstates (only two flavors are considered for simplicity) as:
\begin{equation*}
|\nu_m>=\alpha|\nu_e> +\beta|\nu_\mu>
\end{equation*}
where $\alpha^2+\beta^2=\cos^2{(\theta)}+\sin^2{(\theta)}=1$.
This can be written:
\begin{equation*}
\left(\begin{array}{c}
|\nu_1>\\
|\nu_2>\\
\end{array}\right)=
\left(\begin{array}{cc}
\cos({\theta}) & \sin({\theta})\\
-\sin({\theta}) & \cos({\theta})
\end{array}\right)
\left(\begin{array}{c}
|\nu_e>\\
|\nu_\mu>\\
\end{array}\right)
\end{equation*}
Thus
\begin{equation*}
\left(\begin{array}{c}
|\nu_e(\vec{x},t)>\\
|\nu_\mu(\vec{x},t)>\\
\end{array}\right)=\exp^{i\vec{p}\cdot \vec{x}}\left(\begin{array}{cc}
\cos({\theta}) & \sin({\theta})\\
-\sin({\theta}) & \cos({\theta})
\end{array}\right) \left(\begin{array}{cc}
\exp^{-iE_1t}&0\\
0& \exp^{-iE_2t}
\end{array}\right)
\left(\begin{array}{cc}
\cos({\theta}) & -\sin({\theta})\\
\sin({\theta}) & \cos({\theta})
\end{array}\right)
\left(\begin{array}{c}
|\nu_e(0)>\\
|\nu_\mu(0)>\\
\end{array}\right)
\end{equation*}
If we start with all electron neutrinos:
\begin{equation*}
\left(\begin{array}{c}
|\nu_e(0)>\\
|\nu_\mu(0)>\\
\end{array}\right)=\left(\begin{array}{c}
1\\
0\\
\end{array}\right)
\end{equation*}
This leads to 
\begin{eqnarray*}
<\nu_\mu(\vec{x},t)| \nu_\mu(\vec{x},t)> &=&\sin^2{(2\theta)}\sin^2{\frac{(E_2-E_1)t}{2}}\\
\end{eqnarray*}
and since 
\begin{eqnarray*}
E_2-E_1 &\approx& \frac{m_2^2-m_1^2}{2p}\\
t &\approx& |\vec{x}| \equiv L\\\
p &\approx& E\\
\end{eqnarray*}
\begin{equation*}
P(\nu_e \rightarrow \nu_\mu)=\sin^2{(2\theta)}\sin^2(\frac{\Delta m^2}{4}\frac{L}{E})
\end{equation*}
\subsection{In  $\mathbb{H}$}
Let's examine the time evolution of the electron neutrino, $|e>$, as given in equation ($\ref{eqn:Hcomp}$). 
\begin{eqnarray*}
|e(\vec{x},t)>&=& \exp^{\hat{q}((\mathbf{p} \cdot \mathbf{r} -E t)}|e(0,0)>\\
(\hat{q} &=&\hat{a}i + \hat{b}j +\hat{c}k \mbox{  is a unit imaginary quaternion.})\\
\end{eqnarray*}\\
Thus
\begin{equation*}
|e(\vec{x},t)>=\frac{1}{\sqrt{2}}\bigg(\cos{(\mathbf{p} \cdot \mathbf{r} -E t)} + (\hat{a}i + \hat{b}j +\hat{c}k)\sin{(\mathbf{p} \cdot \mathbf{r} -E t)}\bigg)
\left( \begin{array}{c}
 -\frac{p_x}{E}\cos{(\mathbf{p} \cdot \mathbf{r} -E t)}-\frac{(cp_y-bp_z)}{E}\sin{(\mathbf{p} \cdot \mathbf{r} -E t)} \\
 \\
 a\sin{(\mathbf{p} \cdot \mathbf{r} -E t)}
\end{array}\right) 
\end{equation*}
and
\begin{multline*}
<\mu  (\vec{x},t)| \circ |e  (\vec{x},t)>= \frac{\hat{b}}{2}
\sin{(\mathbf{p} \cdot \mathbf{r} -E t)}\\
 \bigg[\bigg(
\frac{p_y}{E}\cos{(\mathbf{p} \cdot \mathbf{r} -E t)}+\frac{(ap_z-cp_x)}{E}\sin{(\mathbf{p} \cdot \mathbf{r} -E t)}\bigg)
\bigg( \frac{p_x}{E}\cos{(\mathbf{p} \cdot \mathbf{r} -E t)}+\frac{(cp_y-bp_z)}{E}\sin{(\mathbf{p} \cdot \mathbf{r} -E t)}\bigg) \\
+ab\sin^2{(\mathbf{p} \cdot \mathbf{r} -E t)}\bigg]\\
\end{multline*}
Clearly showing that even though, at  $t=0, \quad <\mu| \circ |e>=0$ the $\mu$ neutrino appears as time evolves.
\section{Sterile Neutrinos}
The $|r>$ neutrinos from equation ($\ref{eqn:Hcomp}$) have no lepton flavor and are tentatively identified as "sterile neutrinos," however they have the same helicity as the corresponding flavored neutrinos. 
\section{Conclusions}
The primary objective, showing that a formalism does exist wherein neutrinos can oscillate in flavor without having mass, has been demonstrated.   The neutrinos in this model are spin $\frac{1}{2}$ and have only one helicity. This model also allows for a flavorless fourth type of neutrino. 

The choice for the magnitudes of the three imaginaries, e.g., $a,b,c$ seems to be arbitrary as long as the consistency with $a^2+b^2+c^2=1$ is maintained.  The ubiquity of the combination 
$(\mathbf{\epsilon} \cdot \mathbf{p})$ fuels speculation that to be Lorentz invariant, the $i,j,k$ have to be on equal footing and thus $a=b=c=\frac{1}{\sqrt{3}}$.
\newpage
\renewcommand{\theequation}{A.\arabic{equation}}
\setcounter{equation}{0}  
\section*{Appendix A:  Dirac's Matrices}
\vspace*{.2in}
The representation given by Schiff \cite{Schiff.55} is: 
\vspace*{.3in}
In $\mathbb{C}$
\newline
\vspace*{.3in}
$           
\alpha_x =
\left( \begin{array}{cccc}
 0 & 0 & 0 & 1\\
 0 & 0 & 1 & 0\\
 0 & 1 & 0 & 0\\
 1 & 0 & 0 & 0
      \end{array} \right) 
$
\hspace{.5in}
$\alpha_y =
\left( \begin{array}{cccc}
0 & 0 & 0 & -i\\
0 & 0 & i & 0\\
0 & -i & 0 & 0\\
i & 0 & 0 & 0
      \end{array} \right)
$
\hspace{.5in}
$\alpha_z = 
\left( \begin{array}{cccc}
0 & 0 & 1 & 0\\
0 & 0 & 0 & -1\\
1 & 0 & 0 & 0\\
0 & -1 & 0 & 0
       \end{array} \right)
$    
\vspace{.3in}
\newline
and in  $\mathbb{H}$,  as used by Rotelli \cite{rotelli}    
\vspace{.3in}
\newline
$           
\alpha_x =
\left( \begin{array}{cc}
 0 & i\\
 -i & 0
\end{array} \right) 
$
\hspace{.5in}
$           
\alpha_y =
\left( \begin{array}{cc}
 0 & j\\
 -j & 0
\end{array} \right) 
$
\hspace{.5in}
$           
\alpha_z =
\left( \begin{array}{cc}
 0 & k\\
 -k & 0
\end{array} \right) 
$
\vspace{.3in}
\newpage
\renewcommand{\theequation}{B.\arabic{equation}}
\setcounter{equation}{0}  
\section*{Appendix B: Quaternions}
Quaternions, $\mathbb{H}$, are one of only three ($\mathbb{R}$, $\mathbb{C}$ and $\mathbb{H}$) finite-dimensional division rings\footnote{A field is a commutative division ring. A ring is an algebraic structure which generalizes the algebraic properties of the integers and contains two operations usually called addition and multiplication. One example of a field is the familiar complex numbers. A division ring allows for division (except by zero).  Every field is a ring but non-commutative rings are not fields.} containing the real numbers  $\mathbb{R}$ as a subring - a requirement to preserve probability in quantum mechanics. Quaternion quantum mechanics has been extensively studied and Adler \cite{Adler} has written the definitive reference. $\mathbb{H}$ can be loosely viewed as a non-commutative extension of $\mathbb{C}$.  The imaginary quaternion units, i, j, k are defined by
\[ii = jj = kk = -1\] 
\[ij = -ji = k,\hspace{.3in}ki = -ik = j,\hspace{.3in}jk = - kj = i\]
A general quaternion \emph{q} can be written
\[ q = r + ai + bj + ck\]
where \[\emph{r,a,b,c} \in \mathbb{R}.\]
Every non-zero quaternion has an inverse. 
\begin{eqnarray*} 
q^{-1}&=& \frac{q^*}{(r^2+a^2+b^2+c^2)}\\
\mbox{Therefore  }q^*q&=&(r^2+a^2+b^2+c^2) \in \mathbb{R}\\
\end{eqnarray*}
Quaternion addition is associative: $q_1 + (q_2 + q_3) = (q_1 + q_2) + q_3$ - and defined as 
\[q_1 + q_2 = r_1 + r_2 +(a_1 + a_2)i + (b_1 +b_2)j + (c_1 +c_2)k\]
and quaternion multiplication (paying heed to the non-commutative nature of the imaginary units) is
\begin{eqnarray*}
q_1q_2 &=& (r_1r_2-a_1a_2 - b_1b_2 - c_1c_2) \\
&+&(r_1a_2 + b_1c_2 + b_1c_2 - c_1b_2)i \\
&+&(r_1b_2 - a_1c_2 + b_1r_2 + c_1a_2)j \\
&+&(r_1c_2 + a_1b_2 - b_1a_2 +c_1r_2)k \\
\end{eqnarray*}
Quaternions are associative under multiplication  $(q_1q_2)q_3 = q_1(q_2q_3)$.
A unit imaginary quaternion $q_{\iota}$ is defined as
\[q_{\iota} = a i + b j + c k\]
\begin{center} ( i.e., $r_i=0$ )
\end{center}
where $q_{\iota}^2 = -1$, which means $a^2+ b^2 + c^2  = 1.$\\ \\
It should be noted that many algebraic identities in $\mathbb{C}$  are false in a non-commutative algebra.  For example:
\begin{equation*}
\exp^{i\omega}\exp^{j\omega} \ne \exp^{(i+j)\omega}
\end{equation*}.However Euler's formula is valid:
\begin{equation*}
\exp^{q\omega}=\cos{\omega}+q\sin{\omega}
\end{equation*}
The notation for division
\begin{equation*}
q_3=\frac{q_1}{q_2}
\end{equation*} is ambiguous (on which side of $q_1$ does $q_2^{-1}$ go?) and should not be used.\\ \\
Often useful are the identities:
\[(q_1 q_2)^{-1} = q_2^{-1}q_1^{-1}\]
\[(q_1 q_2)^*=q_2^*q_1^* \]
\noindent It should also be noted that if \emph{i} refers to the \emph{i} of $\mathbb{C}$ rather than of $\mathbb{H}$ it will be specifically indicated. It is also often convenient to represent $i,j,k$ as components of a 3-vector $\mathbf{\epsilon} =(i,j,k)$.\\ \\
\newpage
\section*{Appendix C:  Solving for $u$}
\renewcommand{\theequation}{C.\arabic{equation}}
\setcounter{equation}{0}  
Assuming that $u_1, u_2 \ne$ a function of $x,y,z,t$ and
starting with
\begin{equation*}
\left( \begin{array}{cc}
 \partial_t & -i\partial_x-j\partial_y-k\partial_z\\
 i\partial_x+j\partial_y+k\partial_z & \partial_t\\
\end{array} \right)\\ 
\left( \begin{array}{c}
 u_1\\
 u_2\\
\end{array}\right) \exp^{q(\mathbf{p} \cdot \mathbf{r} -E t)}=0\\
\end{equation*}
leads to \footnote{Frequent use will be made of$(\frac{\mathbf{\epsilon} \cdot  \mathbf{p}}{E})^{-1} = -(\frac{\mathbf{\epsilon} \cdot \mathbf{p}}{E})$ and $ (\frac{\mathbf{\epsilon} \cdot \mathbf{p}}{E}) (\frac{\mathbf{\epsilon} \cdot \mathbf{p}}{E})=-1)$}
\begin{eqnarray}
\left[(u_1\partial_t-u_2(i\partial_x+j\partial_y+k\partial_z)\right]\exp^{q(\mathbf{p} \cdot \mathbf{r} -E t)}&=&0 \label{eqn:u1}\\
\left[u_1(i\partial_x+j\partial_y+k\partial_z)+u_2\partial_t\right]\exp^{q(\mathbf{p} \cdot \mathbf{r} -E t)}&=&0\nonumber \\ \nonumber \\ \nonumber
\end{eqnarray}
These equations have a solution only if the determinant is zero, therefore, (as in $\mathbb{C}$)
\begin{equation*}
E^2=p_x^2+p_y^2+p_z^2
\end{equation*}
and we have both $E_+$ and $E_-$ solutions.  From equation (\ref{eqn:u1})
\begin{eqnarray*}
 -u_1 qE- u_2(\mathbf{\epsilon} \cdot \mathbf{p})q&=&0 \\
-u_1 E- u_2(\mathbf{\epsilon} \cdot \mathbf{p})&=&0\\ 
u_{1+}=u_1=-\frac{u_{2+}}{E_+}(\mathbf{\epsilon} \cdot \mathbf{p})    &   &\\
u_{1-}=\frac{u_{2-}}{|E_-|}(\mathbf{\epsilon} \cdot \mathbf{p})    &   &\\
\end{eqnarray*}
From normalization requirements  
\begin{equation*}
u_1^*u_1+u_2^*u_2=1\\
\end{equation*}
Thus
\begin{equation*}
\frac{(\mathbf{\epsilon} \cdot \mathbf{p})}{E}u_2^*u_2\frac{(-\mathbf{\epsilon} \cdot \mathbf{p})}{E}+u_2^*u_2=1\\
\end{equation*}
and $u_2^*u_2$ must be real and therefore commutes with any quaternion leading to
\begin{equation*}
2u_2^*u_2=1
\end{equation*}
so $u_2=\frac{1}{\sqrt{2}}$ to within an arbitrary phase.
\newpage
\bibliography{Massless}

\begin{thebibliography}{1}

\bibitem{Schiff.55}
Leonard~I. Schiff.
\newblock {\em Quantum Mechanics}.
\newblock McGraw-Hill Book Company, New York, 2nd edition, 1955.

\bibitem{welch1}
Lester~C. Welch.
\newblock Colored and {Q}uaternion {D}irac {P}articles of {C}harges 2/3 and
  -1/3.
\newblock {\em arXiv}, 0809.0484, (2008).

\bibitem{SuK}
Y.~Fukuda et. al.
\newblock Evidence for oscillations of atmospheric neutrinos.
\newblock {\em Phys. Rev. Lett.}, 81, 1998.

\bibitem{cosc}
Dave Casper.
\newblock Neutrino oscillation, mathematical derivation.
\newblock \url{http://www.ps.uci.edu/~superk/oscmath1.html}.
\newblock Accessed: 2016-02-09.

\bibitem{rotelli}
P.~Rotelli.
\newblock The dirac equation on the quaternion field.
\newblock {\em Mod. Phys. Lett}, A4, 1989.

\bibitem{Adler}
Stephen~L. Adler.
\newblock {\em Quaternion Quantum Mechanics and Quantum Fields}.
\newblock Oxford University Press, 1995.

\end{thebibliography}
\bibliographystyle{unsrt}
\end{document}